\documentstyle[preprint,aps,prb,epsfig]{revtex}
\begin{document}
\draft
\title{Can optical spectroscopy directly elucidate the ground state
of $\text{C}_{20}$?}
\author{Alberto Castro,$^{\text{a})}$ 
        Miguel A. L. Marques,$^{\text{b})}$ 
        Julio A. Alonso,$^{\text{a})}$ 
        George F. Bertsch,$^{\text{c})}$
        K. Yabana,$^{\text{d})}$
        and Angel Rubio$^{\text{b})}$}
\address{$^{\text{a})}$ Departamento de F\'{\i}sica Te\'{o}rica, Universidad de
         Valladolid, E-47011 Valladolid, Spain\\
         $^{\text{b})}$ Departamento de F\'{\i}sica de Materiales, Facultad de Qu\'{\i}micas,
         Universidad del Pa\'{\i}s Vasco;\\
         Centro Mixto CSIC-UPV/EHU and
         Donostia International Physics Center (DIPC),
         20080 San Sebasti\'{a}n, Spain\\
         $^{\text{c})}$ Physics Department and Institute for Nuclear Theory,
         University of Washington, Seattle WA 98195 USA\\
         $^{\text{d})}$ Institute of Physics, University of Tsukuba,
         Tsukuba 305-8571, Japan}
\maketitle
\vspace{2cm}
(Received
\tighten
\begin{abstract}
The optical response of the lowest energy isomers of the $\text{C}_{20}$ 
family is calculated using time-dependent density functional theory
within a real-space, real-time scheme.
Significant
differences are found among the spectra of the different isomers,
and thus we propose optical spectroscopy as a tool
for experimental investigation of the structure of these important clusters.
\end{abstract}
\newpage

Fullerenes are carbon clusters formed by the closing of a 
graphitic sheet; the needed curvature is supplied by the insertion,
among a given number of graphitic hexagons, of twelve pentagons
.\cite{Kroto_94}
Besides its most well-known representative, the nearly spherical $\text{C}_{60}$,
a wide variety of fullerenes has been predicted and experimentally observed. However, as 
we reduce the number of atoms, these structures become more reactive
and unstable:
the pentagons present in fullerene-like geometries, although being its major source
of interest, are a cause of strain, specially if two of them are neighbors.
Small fullerenes, with their high proportion of pentagons, have been actively
sought and studied (see, {\it e.g.} Reference 2). Recent discoveries
include the synthesis and purification of the solid form of $\text{C}_{36},$\cite{Piskoti_98}
and the production of the cage and bowl isomers of $\text{C}_{20},$\cite{prinzbach_etal_00} - 
the vibronic fine structure in photoelectron spectra of the cage has also been
recently calculated,\cite{saito_2001} confirming the previous
experimental assignation.

These medium-sized carbon clusters are predicted to possess a wide variety
of isomers like cages, bowls, planar graphitic structures, rings and linear chains.
The theoretical and experimental study of the different isomers
is important, because it may
help us to better understanding the way fullerenes form. 
Several growth mechanisms have been proposed 
over the past years.\cite{Goroff_96}
In the so-called ``pentagon road'', fullerenes grow by the addition
of small carbon fragments to bowl-like structures. The ``fullerene road''
is similar to the pentagon road, but the addition of the small carbon
fragments is made to closed cage-like clusters. However, neither
cages or bowls are usually seen in experiments.\cite{von_helden_etal_93} The
most common technique for carbon cluster formation is laser
vaporization of graphite and subsequent supersonic expansion into
an inert gas atmosphere. At these high temperatures the preferred
isomers are rings and chains. This led Jarrold and co-workers 
to propose a third path, namely that fullerenes grow
by the coalescence and annealing of medium sized carbon rings.\cite{hunter_94}

The smallest possible fullerene, consisting only of 
12 pentagons with no graphitic hexagons intercalated,
is the $\text{C}_{20}$ cage isomer.\cite{Jarrold_00}
Other low energy structures of $\text{C}_{20}$ include a bowl
(which may be considered a $\text{C}_{60}$ fragment),
several rings, and other closed 3D arrangements.
Prinzbach {\em et al} \cite{prinzbach_etal_00} have recently reported
the production of the cage and bowl members of the family. The smallest fullerene cannot be
expected to form spontaneously, but has been produced from a similarly
shaped precursor 
$\text{C}_{20}\text{H}_{20}$, after replacing
hydrogen with Br. The bowl was produced in the same way,
and photoelectron spectroscopy was used to distinguish unambiguously
between the different species.

It has not been possible to make reliable theoretical predictions of 
the most stable structure of $\text{C}_{20}$.
In fact, different levels of theories favor
different geometries: 
at the Hartree-Fock level, the ring is the ground state, followed by the
bowl and the cage.\cite{Grossman_95}
Density functional theory (DFT) \cite{Kohn_Sham_65} in its local density approximation
(LDA) reverses the order, giving the cage 
the lowest energy structure.\cite{Grossman_95,Jones_97} Another
complication is that entropy effects can affect the relative stability.
Molecular-dynamics simulations with the Car-Parrinello method and the LDA 
show that increasing temperature changes the favored structure 
from cage to bowl, then to the ring.\cite{Brabec_92} Better functionals
are availabe with the generalized gradient approximation (GGA), but their
use does not clarify matters:
the ordering of the isomers depends on the correction
used.\cite{Raghavachari_93,Grossman_95,Jones_97} 
Quantum Monte Carlo (QMC) and coupled cluster (CC) methods have 
also been applied in an attempt to resolve the issue,
yielding bowl-ring-cage ordering using the former method \cite{Grossman_95} and
cage-bowl-ring using the latter.\cite{Bylaska_96}
Furthermore, it seems that the results are sensitive to
the pseudopotential employed.\cite{Bylaska_96} Changing slightly the
pseudopotential cutoff radius may actually reverse the ordering of the isomers.

Thus it is important to find a experimental method to determine
the structure that is sensitive enough to be usable with the available
cluster
beam intensities. In this respect optical spectroscopy 
is a useful tool to characterize  
geometries. Some time ago, Rubio {\it et al} \cite{rubio_etal_96} proposed this technique 
to determine the structure of semiconductor and metal clusters.
In particular, they showed that the optical absorption spectra of
different isomers of Si$_4$ and Si$_8$ are sufficiently different
to distinguish easily between them.  The situation is similar with
respect to carbon structures.  The time-dependent density functional
theory (TDDFT) was found to be quite reliable for determining the energies
of the strong transitions in a variety of carbon structures ranging
from chains\cite{yabana_bertsch_97} to conjugated carbon
molecules.\cite{yabana_bertsch_99} In cases where the spectra could be compared with
experiment, the lowest strong transition is typically reproduced to
an accuracy of a few tenth of eV. Comparing rings and chains, the
transition energies differ by several eV, easily allowing 
the structures to be distinguished.
Thus it is quite promising to
use the predicted excitation spectra of C$_{20}$ structures
to distinguish between them.  In this spirit, the present study
is aimed at the
calculation of the optical response of six members of
the $\text{C}_{20}$ family (see Fig.\ 1):
the smallest fullerene (``cage''), which is a Jahn-Teller
distortion of a dodecahedron, the ring, the bowl, and 
three cage-like structures, (d), (e) and (f). The structures
(d) and (f), related by the Stone-Wales transformation,
are quite regular, and composed of four hexagons,
four pentagons, and four four-membered rings.
These clusters are the six members with lower energy as 
calculated by Jones\cite{Jones_99} within the LDA approximation.
Other structures, such as bicyclic rings and chains, may be 
favoured by entropy at high temperature and have been
observed experimentally. However, neither of them seem to 
be a possible ground state.
We will show that calculations based on
time-dependent density functional theory (TDDFT) 
predict characteristic optical spectra
for the ring, cage and bowl species.

We now describe briefly the methodology of the calculation of
optical absorption in the TDDFT.  We start with ground state structures
and electron orbitals, determined with some implementation of DFT.  This
gives the initial condition for solving the time-dependent Kohn-Sham
equation.  Mathematically, there are several very different methods
for solving the equation, but in principle the results should be the same
if the energy functional is the same.  Our method, solving the
equation in real time and representing the wave function on a uniform
spatial grid, is based on a nuclear physics 
algorithm,\cite{flocard_etal_78} and 
has been described several times 
before.\cite{yabana_bertsch_97,yabana_bertsch_99,bertsch}
The real-time response to an impulsive perturbation is Fourier-transformed
to get the dynamic polarizability in the entire range of interest.
Of more direct physical interest is the optical absorption strength
function $S(E)$, obtained
from the imaginary part of the 
polarizability  by the equation
\begin{equation}
S(E)=\frac{2mE}{\hbar^2e^2\pi}\text{Im} \ \alpha (E).
\end{equation}
With this definition, the $f$-sum rule is given by the 
integral $\int S(E) dE = N$.

For the Kohn-Sham energy functional, we used the LDA with the
prescription of Reference 22.  
Use of gradient corrections is possible within this framework, but
results for the optical absorption have been reported to be quite
insensitive to this change.\cite{marques_2001} 
Recent calculations of electronic excitations of a carbon-based molecule
such as benzene \cite{bertsch_schnell_2001,heinze_2000} also show a very close agreement.
Slightly more important modifications
are to be expected if exact-exchange functionals are used, although
the qualitative differences among the spectra should remain.
We also used a pseudopotential to
avoid explicit consideration of the $1s$ electrons, choosing the
norm-conserving soft-core pseudopotential of Reference 26. 
The numerical parameters that need to be specified for the
calculation are: mesh spacing, $0.25$~\AA;
wave function domain, sphere of radius $8$~\AA~ 
(slightly larger in the case of the ring);
time step, 0.001~$\hbar\:\text{eV}^{-1}$; number of time steps,
20,000.  Thus the total propagation time is
$T=20$~$\hbar\:\text{eV}^{-1}$.  One technical point that should be
mentioned is that the Fourier transform over the finite interval
T gives peaks that are broadened by the time cutoff.  In presenting
the results, one removes the spurious oscillations associated with
the time cutoff by multiplying by a filter function, amounting
to a convolution in the frequency domain.  The sum rule is preserved
providing the filter function has zero slope at $t=0$.  In
any case, with $T=20$~$\hbar\:\text{eV}^{-1}$, the 
individual states have a width of about 0.4~eV.

For the calculations reported here, we used structures from two sources.
The geometries of the bowl, cage and ring isomers were determined
by Raghavachari {\em et al}. \cite{Raghavachari_93}
For the (d), (e) and (f) isomers, we used those obtained
by Jones from an all-electron density functional LSD calculation
with an extended Gaussian basis set.\cite{Jones_99} 
As mentioned earlier, the energy differences between the isomers are quite
sensitive to details of the energy functional and the 
pseudopotential.  Fortunately that is not the case at all for the
optical spectra. The optical response is quite
insensitive to changes in the pseudopotential and in the 
energy functional, providing the structures do not change
significantly.  The optical spectra depend very much on the
Kohn-Sham potential, but the differences in that are slight between
the different parameterizations of DFT. The ionization potential (IP)
and electron affinities of the different structures is given in 
Table I.  These quantities are calculated in the DFT by differences
of total energies of systems with differing numbers of electrons but
the same geometry.  The results for the electronic affinity are good to within
0.2~eV if compared to the experimental values reported
by Prinzbach {\em et al} in Reference 4, which
have been obtained through photoelectron spectra.

The results of our TDDFT calculations of the optical absorption
are shown in Table II and Fig. 2.  Table II gives the energies and
strengths of the lowest transitions with appreciable strength, and
Fig. 2 displays the strength function for energies up to the
vacuum ultraviolet.  The solid lines show the TDDFT results 
averaged over all orientations
of the system.  In the top panel we also show by a dashed line
the single-electron response, which corresponds to 
difference of eigenvalues of the HOMO and LUMO orbitals.
The dotted lines in the panels for the ring and bowl show 
the response perpendicular
to the plane of the ring or the bowl center.  This 
direction does not excite $\pi-\pi^*$
transitions in the ring and is almost
negligible in the near ultraviolet frequencies 
(below 8eV),
compared with the response within the plane. 
In the case of the bowl, the perpendicular response cannot be seen in 
the graph because of its extreme weakness in that energy range.
This can be understood because there are no
collective oscillations of the electrons in that direction.
While present molecular beam experiments
are not able to discriminate between the different spatial directions,
the averaged spectra
are still sufficiently different to discriminate between 
the different structures without ambiguity.

We can distinguish two regions in all the graphs: the peaks which 
can be seen in the near ultraviolet, and a broad absorption
that starts at around 7.5~eV. The excitations responsible for this 
latter region are above the ionization threshold,
which range from 7.5 to about 9~eV for all cases considered,
as can be seen from Table I. Since the LDA is unreliable for
describing the ionization process (due to incorrect asymptotic
potential), we will focus
our attention on the relative positioning and intensity of the
lower energy peaks.

The ring exhibits the largest gap in the spectrum and has 
has the strongest collective transition.
The bowl also has a high
gap, larger than 5 eV, but the first significant transition is
an order of magnitude weaker than in the ring.  
The
relative intensities of the peaks, the fact that the first excitation
is divided into two for the bowl, and the relative strength of the
excitation in the 6-7 eV region, 
can all be used to distinguish the bowl from the ring isomer.

The spectra of the four three-dimensional isomers start at much lower
energy and are more similar to 
each other, which is expected from their similar
geometries. The cage isomer shows two clear peaks
at 3.9 and 5.1~eV, with the second much stronger
than the first one. Most of the strength concentrates above the
ionization threshold, and has a broad plateau
starting at around 7~eV. This is clearly
different from planar-like isomers, where an important fraction of
the strength appears below 7~eV.  Isomer (d) can be distinguished
by the presence of a transition at quite low energy, 2.5~eV, as
well as by the fragmentation into many states going up to 6 eV.
Isomer (e) differs from the cage by the presence of a transition
("B") between the transitions that would be seen in the cage.  
The spectrum of isomer (f) is similar to the cage up to the
second peak, but shifted
down by about 0.3~eV.  This is close to the borderline where the TDDFT 
energies are reliable. However, isomer
(f) also has a third peak near 6~eV, in a region where there is
a gap in the cage spectrum, and that difference would be definitive.

We also report calculations for the static polarizability in Table
I.  These have been calculated in two ways, by the formula
\begin{equation}
\alpha(0) =\frac{e^2\hbar^2}{m}
\int_{0}^{\infty} dE  \frac{S(E)}{E^2}, 
\end{equation}
and by adding a static field to the DFT calculation.  The two
methods agree within 2\%, providing an additional check on our
numerical TDDFT computations.  We see that the predicted polarizabilities
differ substantially between ring, bowl, and closed structure, but
there is little discrimination among the closed structures.

In conclusion, we have found that the optical absorption spectra 
calculated in the TDDFT for different candidate structures of 
C$_{20}$ show marked differences that could be used for structural
determination.  Some differences appear already in the visible and
near ultraviolet, and complete discrimination should be possible
with a measurement of the spectrum 
extending up to the 6 eV region.

This work was partially supported by the RTN program of
the European Union NANOPHASE (contract HPRN-CT-2000-00167),
DGESIC (PB98-0345)
and JCyL (VA28/99).
A.C. aknowledges support from the MEC under the graduate
fellowship program, and hospitality from the INT at the
University of Washington. G.B. acknowledges support by the U.S.
Department of Energy under Contract No. E-FG-06-90ER-41132.

\begin{table}
\begin{tabular}{l|l|l|l}
       & I.P. (eV) & E.A. (eV) & $\alpha$ (\AA$^3$) \\     \hline   \hline
ring   &   7.8    &    2.6  (2.44 $\pm 0.03$) & 51              \\ \hline
bowl   &   9.2    &    2.3  (2.17 $\pm 0.03$) & 32              \\ \hline
cage   &   7.5    &    2.1  (2.25 $\pm 0.03$) & 27              \\ \hline
d      &   8.4    &    1.8                    & 28              \\ \hline
e      &   8.0    &    2.8                    & 28              \\ \hline
f      &   7.9    &    3.0                    & 28                 
\end{tabular}
\caption{Ionization potential (I.P.), electron affinity (E.A.) 
and static dipole
polarizability ($\alpha$) of the C$_{20}$ isomers. Experimental
values of the electron affinity from Reference 4 are given between
parentheses.\label{T1}}
\end{table}

\newpage

\begin{table}
\begin{tabular}{l|l|l|l|l|l|l}
        & ring & bowl       & cage       & (d)          & (e)        & (f)        \\ \hline \hline
A & 5.20 (5.4) & 5.05 (0.7) & 3.88 (0.2) &  2.47 (0.03) & 3.77 (0.1) & 3.53 (0.1) \\ \hline
B & 6.42 (1.4) & 5.35 (0.7) & 5.07 (1.3) &  3.23 (0.1)  & 4.33 (0.2) & 4.84 (0.7) \\ \hline
C & 7.09 (2.0) & 6.60 (0.7) &            &  4.21 (0.3)  & 4.96 (0.5) & 5.89 (0.3) \\ \hline
D &            & 7.41 (2.3) &            &  4.67 (0.4)  &            &            \\ \hline
E &            &            &            &  5.86 (0.4)  &            &            \\
\end{tabular}
\caption{Frequencies of selected peaks in the optical response of the studied
responses, in eV. Between parentheses, estimations of their strengths.
\label{T2}}
\end{table}

\newpage
\thispagestyle{empty}
\centerline{Figure 1, Castro et al, Journal of Chemical Physics (communication)}
\vspace{5cm}
\centerline{\hbox{\psfig{figure=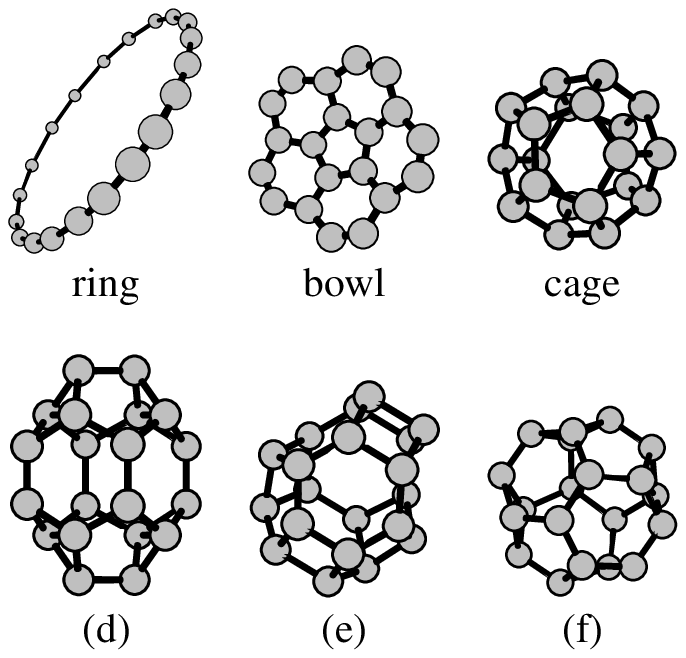,width=8.5cm}}}
\newpage
\thispagestyle{empty}
\centerline{Figure 2, Castro et al, Journal of Chemical Physics (communication)}
\vspace{3cm}
\centerline{\hbox{\psfig{figure=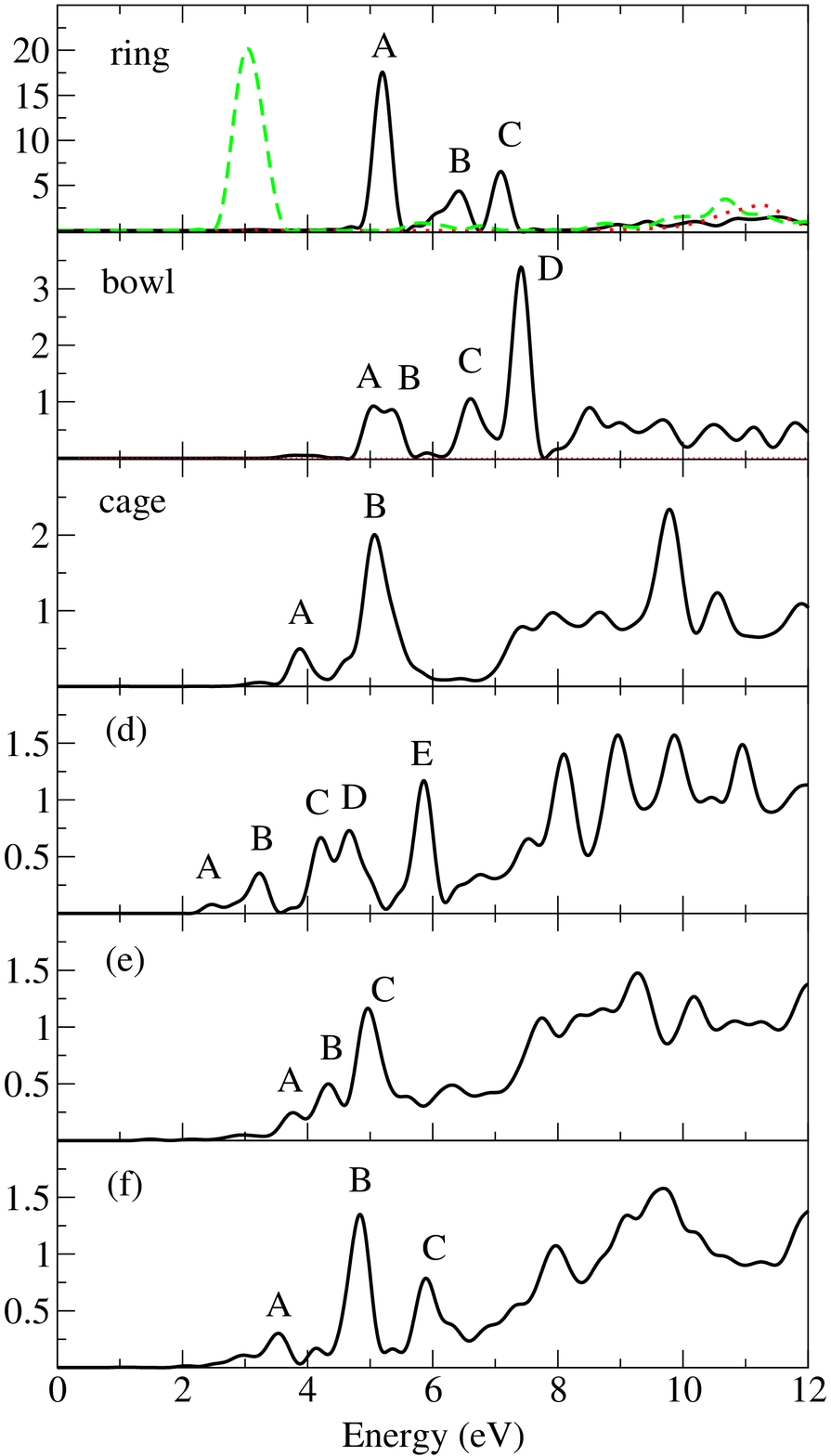,width=8.5cm}}}

\newpage
\thispagestyle{empty}
\vspace*{3cm}
\centerline{FIGURE CAPTIONS}

\vspace{3cm}
Figure 1: {Isomers of C$_{20}$.}

\vspace{2cm}

Figure 2: {Dipole strength function for the isomers of Fig. 1, in 
$\text{eV}^{-1}$, is shown by the solid line. The dashed line
in the upper panel is 
the response in the independent particle
approximation.




\end{document}